\newif\ifnordsec
	
\documentclass{llncs}

\usepackage{graphicx}
\usepackage{xspace}
\usepackage{url}
\usepackage{textcomp}
\usepackage{makeidx}  
\usepackage{subfigure}
\usepackage{tabulary}
\usepackage{booktabs}
\usepackage{hyperref}
\usepackage{multirow}

\begin{document}
\title{Security of OS-level virtualization technologies}

\author{ Elena Reshetova\inst{1},	Janne Karhunen\inst{2},	Thomas Nyman\inst{3},	N. Asokan\inst{4} }

\institute{
Intel OTC, Finland\\
\and
Ericsson, Finland\\
\and
University of Helsinki, Finland\\
\and
Aalto University and University of Helsinki, Finland\\
}

\maketitle

\begin{abstract}
The need for flexible, low-overhead virtualization is evident on many fronts ranging from high-density cloud servers to mobile devices. During the past decade \textit{OS-level virtualization} has emerged as a new, efficient approach for virtualization, with implementations in multiple different Unix-based systems. Despite its popularity, there has been no systematic study of  OS-level virtualization from the point of view of security. 
\ifnordsec
In this paper, we conduct a comparative study of several OS-level virtualization systems, discuss their security and identify some gaps in current solutions.
\else
In this report, we conduct a comparative study of several OS-level virtualization systems, discuss their security and identify some gaps in current solutions.
\fi
\end{abstract}

\section{Introduction}

During the past couple of decades the use of different virtualization technologies has been on a steady rise. Since IBM CP-40~\cite{creasy1981origin}, the first virtual machine prototype in 1966, many different types of virtualization and their uses have been actively explored both by the research community and by the industry. A relatively recent approach, which is becoming increasingly popular due to its light-weight nature, is \textit{Operating System-Level Virtualization}, where a number of distinct user space instances, often referred to as \textit{containers}, are run on top of a shared operating system kernel. A fundamental difference between OS-level virtualization and more established competitors, such as Xen hypervisor~\cite{barham2003xen}, VMWare~\cite{rosenblum1999vmware} and Linux \textit{Kernel Virtual Machine}~\cite{kivity2007kvm} (KVM), is that in OS-level virtualization, the virtualized artifacts are global kernel resources, as opposed to hardware. This allows multiple virtual environments to share a common host kernel and utilize underlying OS interfaces. As a result, OS-level virtualization incurs less CPU, memory and networking overhead, which is important not only for \textit{High Performance Computing} (HPC), such as dense cloud configurations, but also for resource constrained environments such as mobile and embedded devices. The main disadvantage of OS-level virtualization is that each container can only contain a system of the same type as the host environment, e.g. Linux guests on a Linux host.

An important factor to take into account in the evaluation of the effectiveness of any virtualization technology is the level of \textit{isolation} it provides. In the context of OS-level virtualization isolation can be defined as separation between containers, as well as the separation between containers and the host. In order to systematically compare the level of isolation provided by different OS-level virtualization solutions, one first needs to establish a common system model.

The goal of this study is to propose a generic model for a typical OS-level virtualization setup, identify its security requirements, and compare a selection of OS-level virtualization solutions with respect to this model. While other technologies as HW supported secure storage, various encryption primitives and specific CPU/memory features can enhance the security of OS-level virtualization solutions, they are left out of the scope of this paper and present the potential future work. To the best of our knowledge this is the first study of this kind that focuses on the security aspects of OS-level virtualization technologies. We base our analysis on information collected from the documentation and/or wherever possible the source code of the respective systems. As a result of this comparison section~\ref{sec:discussion} identifies a number of gaps in the current implementation of Linux OS-level virtualization solutions. 

\ifnordsec

\else      
			
\section{Usage Scenarios}
\label{sec:usage-scenarios}

We identify the following common usage scenarios as motivation for OS-level virtualization in general. The first three originate from use cases in the context of warehouse scale computing. The latter two stem from security needs.

In \textbf{Server consolidation}, a set of distinct physical servers are substituted with a single physical server running a number of distinct virtual environments. Solutions based on hardware virtualization often require that the guest OS be modified; either to support the virtualization solution itself (as in the case of paravirtualization) or to facilitate interaction between the guest and host OSs by installing special-purpose components into the guest (as in full virtualization solutions such as VMWare, Virtual Box etc.)~\cite{kolyshkin2006virtualization}. In contrast, one of the goals for OS-level virtualization is to provide a set of tools integrated into the OS to allow the creation and management of virtual environment without modifications to the software components placed inside a container. In \textit{Virtual Private Server} (VPS) and cloud computing environments, where service providers grant superuser-level access in the rented virtual environments to customers, strict isolation between environments of different customers and the hosting provider is important unlike in server consolidation where all virtual environments are managed by the same entity.

\textbf{Resource and application state management} emerged from the need to run a number of distinct applications or multiple instances of a single application which require access to the same resources on a single machine, e.g.\ binding to the same network port. In addition by placing an application into a self-contained compartment it is possible to provide \textit{Checkpoint and Restart} (CR) functionality~\cite{biederman2006multiple,kolyshkin2006virtualization,mirkin2008containers}. CR allows processes to be moved between different physical or virtual environments. This can be useful for load-balancing or in high-availability environments, as well as software development and testing on different UNIX platforms. 

A \textbf{Multi-OS experience} allows end-users the ability to use applications and services from different operating systems on the same device by the means of virtualization technology. While the OS-level
virtualization is limited to systems sharing a common kernel, it can provide a way for the user to run a number of different OS variants on the same system. Since there are many new mobile operating
systems on the rise such as Android, Tizen, FirefoxOS and the like, feature is likely to be useful for many experienced users. The need to share certain data, like the user's contacts or calendar, across the different OSs installed on the same device brings in an additional challenge for this use case. 

\textbf{Application or service isolation} places critical and externally exposed services into separate sandbox environments that are able to contain damage in case sandboxed services become compromised. Sandboxing also makes it possible to delegate the administration of these services to third, possibly less trusted, parties~\cite{kamp2000jails}.

The \textbf{Bring Your Own Device} (BYOD) policy~\cite{shim2013bring}  allows one physical device to be used simultaneously for personal and business needs resulting in a need of rigid separation between these two environments in order to guarantee user privacy while conforming to enterprise policies. Presenting separate environments to the end-user can also improve the usability of the solution~\cite{Cellrox}, compared to domain separation by means of access control mechanisms alone.
\fi

\section{System model}

\begin{figure}[ht]
  \subfigure[System model]{%
    \includegraphics[width=0.5\textwidth]{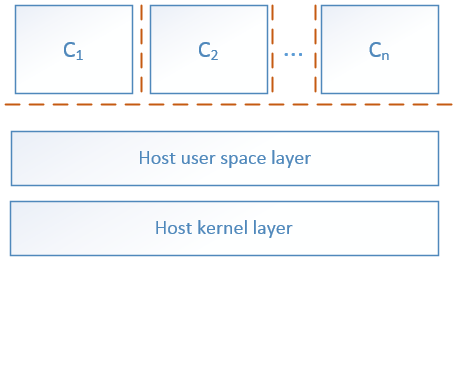} \label{fig:sysmodel}
  }
  \subfigure[Attacker model]{%
    \includegraphics[width=0.5\textwidth]{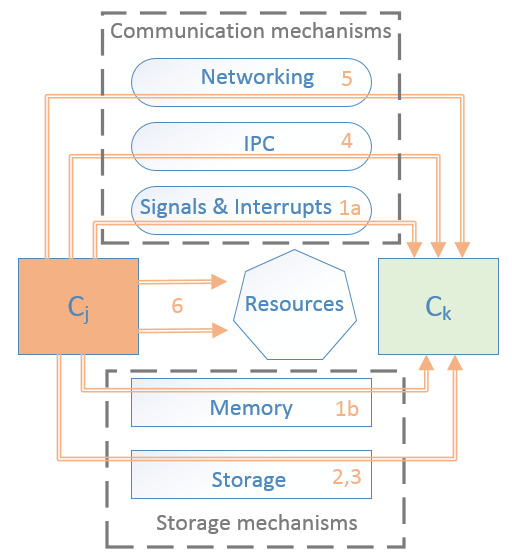} \label{fig:attackermodel}
  }
  \caption{OS-level virtualization}
\end{figure}

\ifnordsec
In Figure~\ref{fig:sysmodel} we present a system model for a typical container setup. 
\else 
In Figure~\ref{fig:sysmodel} we present a system model for a typical container setup that can support the types of usage scenarios we discussed in Section~\ref{sec:usage-scenarios}. 
\fi    
There are a number of containers \ensuremath{\mathit{C_1}}\xspace \ldots \ensuremath{\mathit{C_n}}\xspace that run on a single physical host machine. The OS kernel is shared among all the containers, but the extent of shared host user space depends on a concrete setup (see Table~\ref{tab:model}): 

\textbf{Full OS installation \& management} corresponds to the most common case when the host user space layer comprises a complete OS installation with the container management layer on top. In this case some host resources may be shared between the host and one or more containers via bind-mounts~\cite{bhattiprolu2008virtual} or overlay filesystems~\cite{overlayfs}. Each container can be one of two types:
\begin{itemize}
\ifnordsec
	\item \textit{Application containers} have a single application or service instance running inside. They are most commonly used as sandboxes to contain damage in case an application or a service misbehaves. 
\else 
	\item \textit{Application containers} have a single application or service instance running inside. They are commonly used for application isolation or resource management referred to in Section~\ref{sec:usage-scenarios}.
\fi   
\ifnordsec
	 \item \textit{System containers} have an entire OS user space installation and are commonly used for server consolidation, where a set of distinct physical servers are substituted with a single physical server running a number of distinct virtual environments.  
\else 
		\item \textit{System containers} have an entire OS user space installation and are commonly used for server consolidation. 
\fi  
\end{itemize}

\textbf{Lightweight management} corresponds to the case where the host user space layer consists of merely a light-weight management layer used to initialize and run containers. This setup can be argued to be more secure, as it exhibits a reduced attack surface compared to a complete underlying host system. Again, each container can be one of two types:

\begin{itemize}
	\item \textit{Direct application/service setup} refers to the case when only a single application or service is installed in the container. It is more suitable for application isolation scenarios in which, for instance, a banking application is run in a separate container isolated from the rest of a less trusted OS running in another container.  
\ifnordsec
		\item \textit{Direct OS setup} refers to the case when a container runs an entire OS user space installation. It can provide an end-user the appearance of simultaneously running multiple OS instances, and is therefore well suited for the multi-OS experience that allows end-users the ability to use applications and services from different OS variants on the same device.  
\else 
		\item \textit{Direct OS setup} refers to the case when a container runs an entire OS user space installation. It can provide an end-user the appearance of simultaneously running multiple OS instances, and  is therefore well suited for Multi-OS and BYOD environments. 
\fi
\end{itemize}

\begin{table}[ht]
    \footnotesize
    \centering
    \begin{tabulary}{\linewidth}{C|C|C}
		\toprule
    ~                            												& \multicolumn{2}{c}{\textbf{Container}}                   \\
		\cline{2-3}
    \textbf{Host user space layer}         							& \textit{Application/Service}              & \textit{Full OS installation}   \\
		\hline
		\textit{Full OS installation \& management}   	& Application container 										& System container  \\
		\hline
    \textit{Lightweight management} 							& Direct Application/Service setup 					& Direct OS setup \\
		\toprule 
		\end{tabulary}
		\caption{Types of OS virtualization setups}
    \label{tab:model}
\end{table}

\ifnordsec
The system model described above intentionally omits cases where containers \ensuremath{\mathit{C_1}}\xspace \ldots \ensuremath{\mathit{C_n}}\xspace are not independent, but arranged in a hierarchical structure. While some systems, such as FreeBSD jails~\cite{kamp2000jails}, allow such setups, they are rarely used in practice and are therefore left beyond the scope of this paper. 
\else
The system model described above intentionally omits cases where containers \ensuremath{\mathit{C_1}}\xspace \ldots \ensuremath{\mathit{C_n}}\xspace are not independent, but arranged in a hierarchical structure. While some systems, such as FreeBSD jails~\cite{kamp2000jails}, allow such setups, they are rarely used in practice and are therefore left beyond the scope of this report. 
\fi

\section{Attacker model and security requirements}
\label{sec:attackermodel}

The attacker is assumed to have full control over a certain subset $\bar{C}$ of containers. The remaining set $C$ is assumed to be in the control of legitimate users. The goals of the attacker can be classified as follows:
\begin{itemize}
	\item \textbf{Container compromise:} compromise $C_k \in C$ by means of illegitimate data access, \textit{Man-in-the-Middle} (MitM) attacks or by affecting the control flow of instructions executed in $C_k \in C$.
	\item \textbf{Denial of Service:} disturb normal operation of the host or $C_k \in C$. 
	\item \textbf{Privilege escalation:} obtain a privilege not originally granted to a $C_j \in \bar{C}$. 
\end{itemize}
 
The above goals can be achieved via different types of attacks that can be roughly classified into distinct groups based on the interfaces available in a typical Single UNIX Specification compliant OS~\cite{unixspecification4}. These attack groups can be further arranged into two classes based on the type of underlying mechanism: \textit{attacks via communication mechanisms} and \textit{attacks via storage mechanisms} (see Figure~\ref{fig:attackermodel}). From this classification we derive a set of security requirements that each OS-level virtualization solution needs to fulfill. In the description below, numbers in parenthesis refer to arrows in Figure~\ref{fig:attackermodel}.

\textbf{Separation of processes} is a fundamental requirement that aims to isolate processes running in distinct containers to prevent $C_j \in \bar{C}$ from influencing $C_k \in C$ using interfaces provided by the operating system for process management, such as signals and interrupts (1a).  In addition, it might be possible to directly access the memory of a process running in $C_k \in C$ by using special system calls, e.g.\ the \textsl{ptrace()} system call allows a debugger process to attach and monitor the memory of a debugged process (1b).

\textbf{Filesystem isolation} is required in order to prevent illegitimate access to filesystem objects belonging to $C_k \in C$ or the host (2).

\textbf{Device isolation} should protect device drivers shared between different containers and a host. Such drivers present another significant attack vector because they expose interfaces (3) to code running into the kernel space, which may be abused to gain illegitimate data access, escalate privileges or mount other attacks.

\textbf{IPC isolation} is needed in order to prevent $C_j \in \bar{C}$ from accessing or modifying data belonging to $C_k \in C$ being transmitted over different IPC channels (4). Such channels include traditional System V IPC primitives, such as semaphores, shared memory and message queues as well as POSIX message queues.

\textbf{Network isolation} aims to prevent attacks by $C_j \in \bar{C}$ via available network interfaces (5). In particular, an attacker can attempt to eavesdrop on or modify network traffic of the host or $C_j \in \bar{C}$, perform MitM attacks etc. 

\textbf{Resource management} provides a way to limit the amount of resources available to each container depending on the system load. This is needed in order to prevent an attacker from exhausting physical resources available on a device, such as disk space or disk I/O limits, CPU cycles, network bandwidth and memory (6).

\ifnordsec

\else

\section{Terminology}
\label{sec:terminology}

We define the following terms that are going to be used through the remaining of this report and that might be not familiar for a reader without a Unix/Linux background:

\begin{itemize}
	\item \textbf{Kernel space} refers to the part of the virtual memory that is used to run the OS kernel code, modules, drivers and extensions. 
	\item \textbf{User space} denotes the part of the virtual memory that includes the user applications, system processes (daemons) and services.
	\item \textbf{Kernel resource} is a kernel structure referring to shared physical or virtual devices, system resources, such as memory or cpu time, or a set of identifiers used through the kernel. 
	\item \textbf{Superuser/privileged user} is a special user in UNIX-like systems that is allowed to perform privileged system operations. UNIX classical \textsl{root} user is an example of such user.
	\item \textbf{Linux capabilities} is a set of predefined capabilities implemented in the Linux kernel for performing different privileged operations, such as mounting a filesystem or overwriting the system security policies. Such capabilities are often referred as POSIX capabilities. 
	\item \textbf{Filesystem root} denotes the top-most directory in the UNIX filesystem hierarchy as it is visible to running processes. 
	\item \textbf{Upstream/mainline Linux kernel} refers to the Linux kernel source code tree maintained at \textsl{kernel.org}. This is the official Linux kernel source that contains all the released and upcoming features. 
\end{itemize}
 
\fi

\ifnordsec

\else

\section{Overview of technologies}
\label{sec:overview}

\subsection{FreeBSD Jails}

The pioneering notion of Jails was first introduced in FreeBSD 4.0 in 2000~\cite{kamp2000jails}. The motivation behind Jails was the need to have separate virtual compartments on a single host, combined with the ability to delegate a subset of the traditional superuser privileges to the root user for each compartment. A number of changes were introduced to the FreeBSD kernel in order to implement Jails. These include the hardening of the \textsl{chroot(2)} system call, basic isolation that would restrict a jailed process from communicating with processes outside the Jail, the ability to limit the visibility of processes via the \textsl{procfs} pseudo filesystem and \textsl{sysctl} interfaces, restrictions on TCP/IP networking, Jail-aware device drivers, and the ability to restrict a root user inside a Jail from performing certain system calls. Later on the ability to have multiple IP-addresses per Jail, more powerful Jail management facilities and the ability to create hierarchical Jails were added. 

\subsection{Linux-VServer}

The need for a mechanism similar to FreeBSD Jails in Linux led to the Linux-VServer project~\cite{LinuxVServerProject}. The first official release of the Linux-VServer occurred in 2003. In Linux-VServer, separate virtual environments are referred to as Virtual Private Servers (VPSs). They can be managed with the help of user space tools provided by the \textsl{util-vserver} package. Each VPS has its own context that contains all the information regarding the VPS; its name, allowed limits, bounded capabilities, scheduler information etc. In addition, the behavior of each VPS can be further adjusted by specifying context capabilities and flags that allow a VPS to modify its host name or hide network interfaces that a certain VPS is not permitted to access. The biggest downside with regards to deployability is the need to apply the Linux-VServer patches to kernel source code and recompile the kernel as the Linux-VServer changes are currently not integrated into the mainline Linux kernel development branch. 

\subsection{Solaris Zones}

Solaris Zones/Containers project~\cite{SolarisZonesPrice} was started in 2004 in order to provide a commercial OS-level virtualization solution. Sun engineers analyzed  FreeBSD Jails and Linux-VServer solutions available at the time and concluded, that while the goals of the projects are similar, the depth of OS integration, quality of administrative tools and overall maturity of the aforementioned projects were not at a level needed to support commercial solutions. In addition, they also wanted to create a set of usable zone management tools that would enable the delegation of zone setup and configuration whenever possible to the administrators of a zone. The isolation provided by Solaris Zones is based on attaching a zone identifier to a process, and using it to restrict the visibility of the process across zone boundaries. The zone identifier is also used to determine process privileges inside non-global zones, System V IPC communication etc. Resource management is implemented using standard mechanisms provided in Solaris, such as entitlements, limits and partitions. When used together, they are able to ensure a minimal level of service, bound resource consumption and even the dedication of certain resources only to specific zones.

\subsection{OpenVZ}

The OpenVZ project~\cite{OpenVZProject} is another open source OS-level virtualization solution for Linux begun in 2005. It is currently part of the commercial Parallels Cloud Server solution~\cite{Parallels}. OpenVZ uses the term Virtual Environments (VEs)~\cite{kolyshkin2006virtualization} to refer to containers. The implementation consists of set of kernel changes and user space tools. The OpenVZ kernel is based on the Red Hat Enterprise Linux kernel, which in turn is based on the relatively old 2.6.32 upstream Linux kernel. However, OpenVZ developers have integrated many of their kernel modifications into the upstream kernel. Hence, the project's main user space tool, \textsl{vzctl}, can be used with both upstream and OpenVZ kernels, but the developers ``recommend using the OpenVZ kernel for security, stability and features''.
The OpenVZ project was also the first one to implement the \textit{Checkpoint and Restart} (CR) functionality for for VEs~\cite{mirkin2008containers}. CR allows processes to be moved between different physical or virtual environments. This can be useful for load-balancing or in high-availability environments, as well as software development and testing on different UNIX platforms. 
 
\subsection{LxC}

The Linux Containers (LxC) project~\cite{LxCProject} is the only currently available OS-level virtualization solution for Linux that consists only of a set of user space tools. This is possible because LxC utilizes only those virtualization features integrated into the upstream Linux kernel. This gives LxC an advantage over other Linux-based projects, because the process of applying patches and compiling a specific version of the Linux kernel can become a non-trivial task even for experienced Linux users. Another differentiating feature of LxC is the ability to use Linux Security Modules (LSMs)~\cite{wright2002linux} to harden a container setup. Apparmor~\cite{AppArmor} and SELinux~\cite{smalley2001implementing} profiles are officially supported, but in principle, other existing LSMs, such as Smack~\cite{Smack} could be used as well. 

\subsection{Cells/Cellrox}

The Cells architecture~\cite{andrus2011cells} and the commercial Cellrox solution~\cite{Cellrox} built on the Cells architecture are the only open source OS-level virtualization solutions for smartphones. The primary design goal behind Cells is to support the \textit{Bring Your Own Device} (BYOD) policy~\cite{shim2013bring} on the Android platform. BYOD allows one physical device to be used simultaneously for personal and business needs resulting in a need of rigid separation between these two environments in order to guarantee user privacy while conforming to enterprise policies. Cells allows a user to have two or more virtual phones on the single Android device, e.g. one for personal use and another, which can be controlled by the user's employer and can contain confidential company data and applications. The user can switch between the virtual phones using a special icon on the Android home screen. Similarly to LxC, Cells utilizes upstream kernel features to isolate virtual phones. However, since Android has some non-standard Linux extensions, the developers of Cells had to implement a number of additional isolation mechanisms. The primary example are the changes done to the Binder driver in order to support IPC isolation on Android.

\fi

\section{Comparison}
\label{sec:comparison}

\ifnordsec
We begin with a brief historical overview of the development OS-level virtualization solutions and then proceed to choose some state-of-the-art examples and compare them with respect to each of the security requirements described in the previous section.

\subsection{Evolution of OS-level virtualization}
\label{subsec:evolution}

The history of OS-level virtualization can be traced back to the usage of the \textsl{chroot()} system call in the Unix systems. While its primary goal has been to limit visibility of a filesystem root for a set of processes, \textsl{chroot()} has never been intended as a security mechanism. Nevertheless, it has often been used as a mechanism to limit filesystem access. Such usage is often referred to as \textit{chroot() jails}. However since the \textsl{chroot()} environment is only limited to a filesystem and can be escaped by the privileged user~\cite{chroot}, it cannot be used as it is in order to build a full OS-level virtualization solution. 

The need for a more robust and complete jail implementation in order to have separate virtual compartments on a single host has motivated the emergence of the \textbf{FreeBSD Jails} project~\cite{kamp2000jails} in 2000. The same need in Linux led to the \textbf{Linux-VServer} project~\cite{LinuxVServerProject} with its first release in 2003. In Linux-VServer, separate virtual environments are referred to as Virtual Private Servers (VPSs). \textbf{Solaris Zones/Containers} project~\cite{SolarisZonesPrice} was started in 2004 in order to provide a commercial OS-level virtualization solution. The \textbf{OpenVZ} project~\cite{OpenVZProject}, another open source OS-level virtualization solution for Linux, began in 2005. OpenVZ uses the term Virtual Environments (VEs)~\cite{kolyshkin2006virtualization} to refer to the containers. Both the Linux-VServer and OpenVZ projects provide their own set of kernel patches that in turn create additional usage difficulties. Notably, the OpenVZ project was the first one to implement the \textit{Checkpoint and Restart} (CR) functionality for VEs~\cite{mirkin2008containers}. CR allows processes to be moved between different physical or virtual environments. This can be useful for cluster load-balancing or in high-availability environments, as well as an utility for software development and testing on different UNIX platforms. The \textbf{Linux Containers (LxC)} project~\cite{LxCProject} is the only currently available OS-level virtualization solution for Linux that consists only of a set of user space tools. This is possible because LxC utilizes only those virtualization features already integrated into the upstream Linux kernel
\ifnordsec\footnote{The term \textit{upstream Linux kernel} refers to the Linux kernel source code tree maintained at \textsl{kernel.org}\else\fi. This is the official Linux kernel source that contains all the released and upcoming features.}. Another differentiating feature of LxC is the ability to use Linux Security Modules (LSMs)~\cite{wright2002linux} to harden a container setup. Apparmor~\cite{AppArmor} and SELinux~\cite{smalley2001implementing} profiles are officially supported, but in principle any LSM such as Smack~\cite{Smack} could be used. The \textbf{Cells} architecture~\cite{andrus2011cells} and the corresponding commercial Cellrox solution~\cite{Cellrox} are the only open source OS-level virtualization solutions developed specifically for smartphones. The primary design goal behind Cells is to support the \textit{Bring Your Own Device} (BYOD) policy~\cite{shim2013bring} on the Android platform. BYOD allows one physical device to be used simultaneously for personal and business needs resulting in a need of rigid separation between these two environments in order to guarantee user privacy while conforming to enterprise policies. Similar to LxC, Cells utilizes upstream kernel features to isolate virtual phones. However, since Android has some non-standard Linux extensions, the developers of Cells had to implement a number of additional isolation mechanisms.
\else

Following~\cite{biederman2006multiple}, we define the notion of a \textit{kernel namespace} as a set of identifiers representing a class of global kernel resources, such as process and user ids, IPC objects or filesystem mounts. The OS-level virtualization in the upstream Linux kernel is based on the usage of different kernel namespaces. The subsections below introduce the relevant namespaces and compare selected OS-level virtualization solutions, highlighted in bold in the previous paragraph, based on the security requirements listed in section~\ref{sec:attackermodel}.  

\fi

\subsection{Separation of processes}
\label{subsec:separation}

\begin{table}[t]
    \footnotesize
    \centering
    \begin{tabulary}{\linewidth}{C|C|C}
    \toprule
    
    &
    \textbf{container structure} &
    \textbf{separate namespaces} \\
      \hline
	
	\emph{pros} & 
	simplicity, convenience & 
	flexibility, incremental introduction of containerization  \\
	\hline

	\emph{cons} & 
	 possible information duplication, less flexibility  & 
	 increased complexity\\
  \hline
	
	\emph{used by} & 
	FreeBSD, SolarisZones, Linux-VServer, OpenVZ & 
	Linux-VServer, OpenVZ, LxC, Cells \\
	\toprule
    
    \end{tabulary}
    \caption{Comparison of containerization approaches}
    \label{tab:comparison:process}
\end{table}

The primary isolation mechanism required from any OS-level virtualization solution is that it is able to distinguish processes running in different containers from those running on the host, limit cross-container process visibility and to prevent memory and signaling-level attacks described in the section~\ref{sec:attackermodel}. The simplest solution to this problem is to embed a container identifier \ensuremath{\mathit{C_i}}\xspace into the process data structure and to check the scope and the permissions of all syscall invocations. 

FreeBSD Jails, Solaris Zones, OpenVZ and Linux-VServer implementations follow this approach by linking a structure describing the container to the process data structure. However, unlike FreeBSD and Solaris, the data structures describing OpenVZ and Linux-VServer containers are not used to achieve process separation. They only store related container data such as resource limits and capabilities. Instead, OpenVZ, Linux-VServer, LxC and Cells use \textit{process id (pid) namespaces} that are part of the mainline Linux kernel. A pid namespace is a mechanism to group processes in order to control their ability to see (for example via \textsl{proc} pseudo-filesystem) and interact (for example by sending signals) with one another. The pid namespaces also provide pid virtualization: two processes in different pid namespaces may have the same pid. 

Having a separate structure describing a container and storing a pointer to it in the process task structure is a convenient way to have all the relevant information concerning the container in one place. However, the upstream Linux kernel has followed a different approach of grouping different kernel resources into separate namespaces and using these namespaces to build containers. This approach incurs additional complexity, but adds the flexibility to choose a combination of namespaces that best fits the desired use case. It also allows gradual introduction of namespaces to an existing system, like the upstream Linux kernel, which also helps in testing and verification of the implementation~\cite{biederman2006multiple}. Furthermore, it avoids information duplication when both the process and the container structures have similar information. The pros and cons of these two approaches are summarized in Table~\ref{tab:comparison:process}.  

In addition to the ability to isolate and virtualize process ids, the upstream Linux kernel also allows virtualization and isolation of the user and group identifiers with the help of \textit{user namespaces}. Typically the root user has all the privileges to perform various system administration tasks and is able to override all access control restrictions. However, it is not desired that a root user running inside a container would be given the privileges of the host root user. Therefore, the Linux user namespace implementation interprets a given Linux capability as authorizing an action within that namespace: for example, the \textsl{CAP\_SYS\_BOOT} capability inside a container grants the authority to reboot that container and not the host. Moreover, many capabilities such as \textsl{CAP\_SYS\_MODULE} cannot be safely granted for container in any meaningful manner. When a process attempts to perform an action guarded by such capability, the kernel always checks if the process possesses this capability in the host user namespace. All Linux OS-level virtualization solutions support the option of starting a new user namespace for each container, but all the related configuration such as mapping the user identifiers between the host and the container must be done manually. 

\subsection{Filesystem isolation}
\label{subsec:filesystem}

The filesystem is one of the most important OS interfaces that allows processes to store and share data as well as to interact with one another. In order to prevent filesystem-based attacks described in section~\ref{sec:attackermodel}, it should be possible to isolate the filesystem between containers and to minimize the sharing of the data. The amount of sharing needed between the host and each container depends on the usage scenario. In the case of application isolation, it is not worthwhile to completely duplicate the OS setup inside a container and therefore some parts of the filesystem, such as common libraries, need to be securely shared with the host. On the other hand in the case of server consolidation, quite often it is best to completely separate the filesystems and create container filesystems from scratch.  

All Linux-based OS virtualization solutions utilize a \textit{mount namespace} that allows separation of mounts between the containers and the host. The design of upstream Linux mount namespaces\cite{biederman2006multiple} has been influenced by private namespaces~\cite{pike1992namespace} in Plan 9 from Bell Labs~\cite{pike1990plan9}. Namespaces in Plan 9 are file-orientated, and the principal purpose is to facilitate the customization of the environment visible to users and processes. Since all Linux based systems create each container within a new mount namespace, all the internal mount events are only effective inside the given container. However, it is important to underline that the mount namespace by itself is not a security measure. Running a container in a separate mount namespace does not give any additional guarantees concerning the data isolation between the containers since containers inherit the view of filesystem mounts from their parent and thus are able to access all parts of the filesystem similarly. 

A typical approach for process filesystem access containment is by using the \textsl{chroot()} system call where process is bound within a subtree of the filesystem hierarchy. If desired, resources may be shared with the host by mounting them within the subtree visible inside the container. Since the \textsl{chroot()} system call~\cite{linuxman} only affects pathname resolution, privileged processes (i.e.\ processes with the \emph{CAP\_SYS\_CHROOT} privilege) can escape the chroot jail. This can be done for example by changing the root directory again via \textsl{chroot()} to a subdirectory relative to their current working directory. Of the virtualization solutions under comparison, only Cells relies on \textsl{chroot()} alone. Some systems, such as Linux-VServer utilize a \textit{Secure chroot barrier}~\cite{LinuxVServerProject} to prevent processes in a VPS from escaping the modified environment. 

Another approach, utilized by for instance LxC, is to not only modify the root directory for processes in a container, but modify the \emph{root filesystem} as well. This can be achieved with the Linux specific \textit{pivot\_root()} system call~\cite{linuxman}, which is typically used during boot to change from a temporary root filesystem (e.g.\ an initrd) to the actual root filesystem. As its name suggests, the \textsl{pivot\_root()} system call moves the mountpoint of the old root filesystem to a directory under the new root filesystem, and puts the new root filesystem at its place. When done inside a mount namespace, the old root filesystem can be unmounted, thus rendering the host root filesystem inaccessible for processes inside the container, without affecting processes belonging to the root mount namespace on the host system. At the time of writing, the implementation of \textsl{pivot\_root()} also changes the root directory and current working directory of the process to the mountpoint of the new root filesystem if they point to the old root directory. OpenVZ relies on this behavior and uses the \textsl{pivot\_root()} system call alone. However, as the behavior with regards to the current root directory and the current working directory remains unspecified, proper usage dictates that the caller of \textsl{pivot\_root()} must ensure that processes with root directory or current working directory at the old root operate correctly regardless of the behavior of \textsl{pivot\_root()}. To ensure this, LxC changes the root directory and current working directory to the mountpoint of the new root before invoking \textsl{pivot\_root()}.

FreeBSD and Solaris also provide a sandbox-like environment for each jail/zone using similar \textsl{chroot()}-like calls that are claimed to avoid above mentioned security vulnerabilities~\cite{kamp2000jails},~\cite{SolarisZonesPrice}. Mounting and unmounting of filesystems is prohibited by default for a process running inside a jail unless different \textsl{allow.mount.*} options are specified. 

A separate user namespace per container can further strengthen the filesystem isolation by mapping the user and group ids to a less privileged range of host uids and groups. Together with a mount namespace and a \textsl{pivot\_root} environment it strengthens protection against filesystem-based attacks described in~\ref{sec:attackermodel}.

\subsection{Device isolation}
\label{subsec:devnodes}

In Unix, device nodes are special files that provide an interface to the host device drivers. In classical Unix configurations, the device nodes are separated from the rest of the filesystem and their inodes are placed in the \textsl{/dev} directory. In the case of Linux, this task is usually performed by the udevd daemon process issuing the \textsl{mknod} system call upon receiving the event from the kernel. Device nodes are security-sensitive since an improperly exposed or shared device inside a container can lead to a number of easy attacks (see section~\ref{sec:attackermodel}). In the simplest example, if a container has an access to \textsl{/dev/kmem} and \textsl{/dev/mem} nodes, it is able to read and write all the memory of the host. Thus, in order to isolate containers from one another it is important to prevent containers from creating new device nodes and to make sure that containers are only allowed to access a ``safe'' set of devices listed below:

\begin{enumerate}
	\item \textbf{Purely virtual devices}, such as pseudo-terminals and virtual network interfaces. The security guarantee comes from the fact that these devices are explicitly created for each container and not shared.
  \item \textbf{Stateless devices}, such as \textsl{random}, \textsl{null} and others. Sharing these devices among all containers and the host is safe because they are stateless. 
  \item \textbf{User namespace-aware devices}. If a device supports verifying process capabilities in the corresponding user namespace, then it is safe to expose such device to a container, because the specified limitations will be enforced. The current 3.14-rc2 upstream kernel does not have any physical devices supporting this feature, but they are expected to appear in the future.   
\end{enumerate}

All compared systems allow the system administrator to define a unique set of device nodes for each container and by default create only a small set of stateless and virtual devices. In Linux, creation of new device nodes within containers can be controlled by limiting access to the \textsl{CAP\_SYS\_MKNOD} Linux capability and by ensuring that all mountpoints inside containers have the \textsl{nodev} flag set. 

The biggest difference of the Cells implementation is the addition of a ``\textit{device namespace}'' that attempts to make the Linux input/output devices namespace-aware. Cells assumes the host to have a single set of input/output devices and multiplexes access to the physical host device via virtual devices created in each container. One virtual device at a time is allowed to access physical devices, based on whether an application from a given container is ``on the foreground'' (ie. visible on the screen) or not. Security-wise such an exclusive-access solution is comparable to the ``purely virtual'' devices category mentioned above and can be considered safe.

As mentioned above, Linux device drivers controlling physical devices are currently not namespace-aware and thus cannot be securely used inside containers. Quite commonly these devices assume only one controlling master host and require privileges that are hard to grant for a unprivileged container securely (unless the device is used exclusively by a single container). In other words, namespace support inside the device drivers would require extensive modifications to the existing driver code base. 

\subsection{IPC isolation}
\label{subsec:ipc}

In order to achieve IPC isolation between containers, processes must be restricted to communicate via certain IPC primitives only within their own container. If the filesystem isolation is done correctly (see section~\ref{subsec:filesystem}), then filesystem-based IPC mechanisms (such as UNIX domain sockets and named pipes) are automatically isolated because the processes are not able to access filesystem paths outside of their own container. However, the isolation of the rest of the IPC objects (such as System V IPC objects and POSIX message queues) requires additional mechanisms. In Linux these IPC objects are isolated with the help of the \textit{IPC namespaces} that allow the creation of a completely disjoint set of IPC objects. Linux-VServer, OpenVZ, LxC and Cells all spawn a new IPC namespace for each container in order to achieve the required isolation. 

In addition to using IPC namespaces, Cells also has to implement namespace support for the Binder system since it is the primary IPC mechanism on the Android OS. The solution~\cite{BinderNamespace} includes having a separate Context Manager for each IPC namespace that is able to resolve Binder addresses only in that namespace and therefore provide isolation of Binder addresses between different containers. 

Solaris Zones follow a different approach to isolate IPC objects that are not filesystem path-based. A zone ID is attached to each object based on the zone ID of the process that creates it, and processes are not able to access objects from other zones. An exception is made only for an administrator in the global zone that can access and manage all the objects. FreeBSD simply blocks SysV IPC object-related system calls if such calls are issued from within a jail. The \textsl{allow.sysvipc} option allows SysV IPC mechanisms for jailed processes but lacks any isolation between jails.  

\subsection{Network isolation}
\label{subsec:network}

\begin{table}[t]
    \footnotesize
    \centering
    \begin{tabulary}{\linewidth}{C|C|C|C}
    \toprule
    
    &
    \textbf{Layer 3 bind filtering} &
    \textbf{Layer 3 VNI} & 
    \textbf{Layer 2 VNI} \\
  \hline
	
  \emph{traffic shaping and policing} & 
	no & 
	yes &
	yes\\
  \hline
	
		\emph{separate routing and filtering tables} & 
	no & 
	no &
	yes\\
  \hline	

		\emph{used by} & 
	FreeBSD Jails, Linux-VServer & 
	Solaris Zones, OpenVZ &
	Solaris Zones, OpenVZ, LxC, Cells \\
	\toprule

    \end{tabulary}
    \caption{Comparison of network isolation}
    \label{tab:comparison:network}
\end{table}

The main goal of network isolation is to prevent network-based attacks described in section~\ref{sec:attackermodel}. 
\ifnordsec
Moreover, in order to be able to support applications that might contend for the same type of network resource (such as binding to the same network port), it also needs to provide a virtualized view of the network stack. 
\else
Moreover, in order to fulfill the server consolidation and resource management use cases, it also needs to provide a virtualized view of the network stack. 
\fi

Network isolation methods differ in terms of the OSI layer of the TCP/IP stack where the isolation is implemented (see Table~\ref{tab:comparison:network} for a comparison between these implementations). FreeBSD and Linux-VServer implement network isolation on Layer 3 with the help of bind filtering. They restrict a \textsl{bind()} call made from within a container to a set of specified IP addresses and therefore processes are only allowed to send and receive packets to/from these addresses. The benefit of such an approach is the small amount of code that needs to be modified in the network implementation and a minimal performance overhead. However, the downside is that a lot of the standard networking functionality is not accessible for a process inside a container such as obtaining an address from the Dynamic Host configuration Protocol (DHCP), acting as a DHCP server or the usage of routing tables.  

Another approach, supported by Solaris Zones and OpenVZ, provides a Layer 3 virtualized network interface (VNI) for each container. Compared to bind filtering this implementation is more flexible since it allows the configuration of different traffic control settings, such as traffic shaping and policing, from within the container. The Layer 3 implementation provided by OpenVZ is called \textsl{venet}, while Solaris uses the term \textsl{shared-IP zone}. 

The third approach includes providing a Layer 2 virtualized network interface for each container with a valid Link layer address. This gives containers the ability to use many features that are not supported by the previous two solutions, such as DHCP autoconfiguration, separate routing information and filtering rules. This approach can also support a broader set of network configurations. However, the primary downsides include a performance penalty and the inability to control the container networking setup from the host. The latter can be important for the server consolidation case if the host administrator needs to be in the control of the overall network configuration. OpenVZ, Solaris, LxC and Cells all support the creation of the Layer 2 virtualized interfaces. On Linux platforms this feature is called virtual Ethernet (\textsl{veth}). On Solaris a similar configuration is named \textsl{exclusive-IP zone}.

The Linux Layer 2 network isolation is based on the concept of a \textit{network namespace} that allows the creation of a number of networking stacks that appear to be completely independent. The simplest networking configuration for a container running in a separate network namespace includes a pair of virtually linked Ethernet (\textsl{veth}) interfaces and assigning one of them to the target namespace while keeping the other one in the host namespace. After the virtual link is established, interfaces can be configured and brought up~\cite{NetworkNSSetup}. 

Linux provides multiple ways for connecting containers to physical networks. One option is connecting the \textsl{veth} interface and the host physical interface by using a virtual network bridge device. Another option is to utilize routing tables to forward the traffic between virtual and physical interfaces. When a virtual bridge device is used, all container and host interfaces are attached to the same link layer bridge and thus receive all link layer traffic on the bridge. However, in the case of route configuration, containers are not able to communicate with each other unless a network route is explicitly provided. Also in the latter case, container addresses are not visible to outsiders like in bridged mode. Another way of providing network connectivity for containers is to use the \textsl{MACVLAN} interface~\cite{LxCProject} that allows each container to have its own separate link layer address. MACVLAN can be set to operate in a number of modes. In a private mode containers cannot communicate with each other or the host making it the strictest isolation setup. The bridge mode allows containers to communicate with one another, but not with the host. The Virtual Ethernet Port Aggregator (\textsl{VEPA}) mode by default isolates containers from one another, but leaves the possibility to have an upstream switch that can be configured to forward packets back to the corresponding interface. Currently LxC is the only solution that can support all the \textsl{MACVLAN} modes. 

\subsection{Resource limiting}
\label{subsec:resources}

\begin{table}[t]
    \footnotesize
    \centering
    \begin{tabulary}{\linewidth}{C|C|C}
    \toprule
    
    &
    \textbf{\textsl{rlimits}} &
    \textbf{\textsl{cgroups}} \\
  \hline
  
  \emph{scope} & 
  per process, inheritable &
  per process group, inheritable \\
  \hline
	
	\emph{managed resources} & 
  memory(limited), CPU(limited), filesystem, number of threads &
  memory, CPU, block I/O, devices, traffic controller  \\
  \hline
	
	\emph{action when limit is reached} & 
  resource request denial and process termination &
  resource request denial, possibility to have a custom action   \\
  \hline
		
	 \emph{used by} & 
    Linux-VServer, Cells &
    OpenVZ, LxC, Linux-VServer, Cells  \\
   \toprule

    \end{tabulary}
    \caption{Comparison of Linux resource management mechanisms}
    \label{tab:comparison:resource}
\end{table}
A good virtualization solution needs to provide support for limiting the amount of primary physical resources allocated to each container in order to prevent containers from carrying out denial of service attacks described in section~\ref{sec:attackermodel}.  

Since the 9.0 release FreeBSD utilizes Hierarchical Resource Limits (RCTL) to provide resource limitation for users, processes or jails~\cite{RCTL}. RCTL supports defining an action in case a specified limit is reached: deny new resource allocation, log a warning, send a signal (for example SIGHUP or SIGKILL) to a process that exceeded the limit or to send a notification to the device state change daemon.

Solaris implements resource management for zones using a number of techniques that can be either applied to a whole zone or to a specific process inside a zone. Resource partitioning, called \textit{resource pools}, allows defining a set of resources, such as a physical processor set, to be exclusively used by a zone. A dynamic resource pool allows to adjusting the pool allocations based on the system load. Resource capping is able to limit the amount of the physical memory used by a zone. 

The traditional way of managing resources on BSD-derived systems is the \textsl{rlimits} mechanism that allows specifying soft and hard limits for system resources for each process. Cells and Linux-VServer utilize \textsl{rlimits} to do resource management for containers. However, the main problem of rlimits is that it does not allow specifying limits for a set of processes or to define an action when a limit is reached. Also the CPU and memory controls are very limited and do not allow specifying the relative share of CPU time, number of virtual pages resident in RAM or physical CPU or memory bank allocations.

In an attempt to address some of these limitations, OpenVZ and Linux-VServer have implemented custom resource management extensions, such as new limits for the maximum size of shared and anonymous memory or new CPU scheduler mechanisms.  In addition both virtualization solutions added the possibility to specify resource limits per container.

\textit{Linux Control Groups (cgroups)}~\cite{cgroupsKernelorg} is a relatively new mechanism that aims to address the downsides of \textsl{rlimits}. It allows arranging a set of processes into hierarchical groups and performs resource management for the whole group. The CPU and memory controls provided by \textsl{cgroups} are rich, and in addition it is possible to implement a complex recovery management in case processes exceed their assigned limits. LxC, Linux-VServer, OpenVZ and Cells provide a way to use cgroups as a container resource management mechanism. 

Table~\ref{tab:comparison:resource} presents a comparison of different aspects between \textsl{rlimits} and \textsl{cgroups}. A combined use of these mechanisms allows protecting the container from a set of DoS attacks directed towards the CPU, memory, disk I/O and filesystem (\textsl{rlimits} combined with \textsl{filesystem quotas}). However, the future direction is to aggregate all resource management to \textsl{cgroups}, and allow \textsl{rlimits} to be changed by a privileged user inside a container\footnote{documentation in source code of \url{http://lxr.linux.no/\#linux+v3.13.5/kernel/sys.c\#L1368}}. 

\section{Related work}
\label{sec:relatedwork}

A number of previous studies have compared different aspects of the OS-level virtualization to other virtualization solutions. Padala et al.~\cite{padala2007performance} analyze the performance of Xen vs. OpenVZ in the context of server consolidation. Chaudhary et al.~\cite{chaudhary2008comparison}, Regola et al.~\cite{regola2010recommendations} and Xavier et al.~\cite{xavier2013performance} perform comparisons of different virtualization technologies for HPC. Yang et al.~\cite{yang2013impacts} study the impact of different virtualization technologies for the performance of the Hadoop framework~\cite{kizza2013virtualization}.

The Capsicum sandboxing framework~\cite{watson2010capsicum} introduced in FreeBSD 9 isolates processes from global kernel resources by disabling system calls which address resources via global namespaces. Instead, resources are accessed via capabilities which extend Unix file descriptors. Linux has a similar mechanism, called \textit{seccomp}~\cite{seccomp}, that allows a process to restrict a set of systems calls that it can execute. Both Capsicum and \textsl{seccomp} require modifications to existing applications.    

\ifnordsec
While there are OS-level virtualization solutions such as ICore~\cite{icore} and Sandboxie~\cite{sandboxie} in existence for Microsoft Windows as add-on solutions, we have left them out this paper's scope due to their closed nature. Authors are not aware of any OS-level virtualization solutions for Mac OS X or iOS.
\else
While there are OS-level virtualization solutions such as ICore~\cite{icore} and Sandboxie~\cite{sandboxie} in existence for Microsoft Windows as add-on solutions, we have left them out this report's scope due to their closed nature. Authors are not aware of any OS-level virtualization solutions for Mac OS X or iOS.
\fi

In addition to the OS-level virtualization solutions under comparison in this study, researchers have developed a number of other technologies. An attempt by Banga et al.~\cite{banga1999resource} to do fine-grained resource management led to the creation of a new facility for resource management in server systems called \textit{Resource Containers}. Zap~\cite{osman2002design} allows the grouping of processes into \textit{Process Domains} (PODs) that provide a virtualized view of the system and support for CR. An OS-level virtual machine architecture for Windows is proposed by Yu et al.~\cite{yu2006feather}. A partial OS-level virtualization is provided by the PDS environment by Alpern et al.~\cite{alpern2005pds}. Wessel et al.~\cite{wessel2013improving} propose a solution for isolating user space instances on Android similar to the Cells/Cellrox. The solution by Wessel et al.\ has a special focus on security extensions, such as remote management, integrity protection and storage encryption. 

\section{Discussion and Conclusions}
\label{sec:discussion}

All compared systems implement core container separation features in terms of the memory, storage, network and process isolation. However, while the initial innovation around containers happened on FreeBSD and Solaris, the mainline Linux has caught up in terms of features and the flexibility of the implementation.  Linux is likely to have a complete user space process environment virtualization in course of time. Given the scale of deployment of Linux and the maturity of its OS-level virtualization features, we focus on Linux in the rest of this section.

Table~\ref{tab:summary} summaries the state of the OS-level virtualization supported by the current upstream Linux kernel. The first row shows how each type of isolation discussed in section~\ref{sec:comparison} can be achieved using the currently available techniques. The second row presents a number of gaps that are briefly described below.

\begin{table}[t]
\begin{tabular}{c|c|c|c|c|c|c}
\hline
                                                                                & \textit{\textbf{\begin{tabular}[c]{@{}c@{}}separation \\ of\\ processes\end{tabular}}} & \textit{\textbf{\begin{tabular}[c]{@{}c@{}}file-\\ system\\ isolation\end{tabular}}} & \textit{\textbf{\begin{tabular}[c]{@{}c@{}}IPC \\ isolation\end{tabular}}} & \textit{\textbf{\begin{tabular}[c]{@{}c@{}}device \\ isolation\end{tabular}}}                      & \textit{\textbf{\begin{tabular}[c]{@{}c@{}}network \\ isolation\end{tabular}}}           & \textit{\textbf{\begin{tabular}[c]{@{}c@{}}resource\\ limiting\end{tabular}}} \\ \hline
\multirow{2}{*}{\textit{\begin{tabular}[c]{@{}c@{}}achieved\\ by\end{tabular}}} & pid ns                                                                                 & \multicolumn{3}{c|}{mount ns, pivot\_root}                                                                                                                                                                                                                             & \multirow{2}{*}{\begin{tabular}[c]{@{}c@{}}network  ns,\\ veth, \\ MACVLAN\end{tabular}} & \multirow{2}{*}{\begin{tabular}[c]{@{}c@{}}rlimits,\\ cgroups\end{tabular}}   \\ \cline{2-5}
                                                                                & \multicolumn{2}{c|}{user ns}                                                                                                                                                  & ipc ns                                                                     & \begin{tabular}[c]{@{}c@{}}cgroups device\\ controller,\\ exclusive\\  device usage\end{tabular}   &                                                                                          &                                                                               \\ \hline
\textit{\begin{tabular}[c]{@{}c@{}}open\\ problems\end{tabular}}                & \multicolumn{2}{c|}{security ns}                                                                                                                                              & \begin{tabular}[c]{@{}c@{}}IPC \\ extensions\end{tabular}                  & \begin{tabular}[c]{@{}c@{}}device ns, \\ (pseudo)random \\ devices,\\ hotplug support\end{tabular} & n/a                                                                                      & \begin{tabular}[c]{@{}c@{}}incomplete\\ cgroups\end{tabular}                  \\ \hline
\end{tabular}
\caption{Summary of OS-level virtualization in upstream Linux kernel}
\label{tab:summary}
\end{table}

\textbf{Security namespaces.} In order to reduce security exposure and adhere to the \textit{principle of least privilege}, many OSs provide an integrated mandatory access control (MAC) mechanism. MACs can be used to strengthen the isolation between different containers and the host, as well as to enforce MAC policies for processes inside containers. The latter is especially important when the container has a full OS installation, because it usually comes with pre-configured MAC policies. Therefore, OS-level virtualization solutions should support the ability to use the common MAC mechanisms in the underlying host kernel to enforce independently defined (container-specific) MAC policies. However, currently none of the compared solutions fulfills this requirement. Linux kernel developers plan to address this limitation in the future by introducing a \textit{security namespace} that would make LSMs container-aware.

\textbf{IPC extensions.} While IPC namespaces and filesystem isolation techniques cover most of the inter-process communication methods available on Linux, exceptions exist. For example \textit{Transparent Inter-process Communication} (TIPC)~\cite{TIPCProject} is not currently covered. TIPC is a network protocol that is designed for an inter-cluster communication. Usage of such methods would break the IPC isolation borders between containers and if the given features are not needed, they should be disabled from the kernel configuration.

\textbf{Device namespaces.} As discussed in section~\ref{subsec:devnodes}, secure access to device drivers from within a container remains an open problem. One way to approach it would be to create a new namespace class (a \textit{device namespace}) and group all devices to belong in their own device namespaces in hierarchical manner, following the generic namespace design pattern. Given this, only processes within the same device namespace would be allowed to access devices belonging in it. However, since the core of such functionality would resemble more access/resource control than a fully featured namespace, it was initially decided to implement the functionality as a separate \textsl{cgroups} device controller. The discussions defining the full notion of the device namespace and its functionality continue in the kernel community~\cite{devicens}.

\textbf{(Pseudo)random number generator devices.} In section~\ref{subsec:devnodes} we stated that using stateless devices such as \textsl{/dev/random} or \textsl{/dev/urandom} are secure within containers due to their stateless nature. This means that even if two containers share the same device, they cannot predict or influence the output from another device node within another container. However, it is important to note that exposing blocking devices, such as \textsl{/dev/random}, poses a Denial-of-Service possibility. A malicious container can exhaust all available entropy and block the \textsl{/dev/random} from being used in all other containers and the host, making it impossible to perform cryptographic operations requiring random input. Even if only non-blocking \textsl{/dev/urandom} is exposed, there is a theoretical possibility that a malicious container can predict the random output for another container or a host. For example in~\cite{dodis2013security} Dodis et al. give an assessment of both \textsl{/dev/random} and \textsl{/dev/urandom} showing that these devices do not accumulate entropy properly. A complete solution would be to implement a separate random device per namespace or even introduce a namespace for (pseudo)random number generators.

\textbf{Hotplug support.} Desktop Linux relies heavily on the dynamic nature of device nodes. Once new devices are plugged in to the system, the kernel generates an \textsl{uevent} structure notifying the user space of the new hardware. As briefly explained in section~\ref{subsec:devnodes}, \textsl{Uevent} is typically handled by the \textsl{udevd} daemon which configures the device for system use. Traditionally it has also created the corresponding device node after device setup. As far as containers are concerned, this setup is risky and complicated - containers should not be allowed to configure hardware and/or have permissions for creating the new device nodes. As a result, safe device hotplug for containers remains an open problem.

\textbf{Incomplete implementation of \textsl{cgroups}.} As was mentioned in the section~\ref{subsec:resources}, the current goal of the upstream Linux is to integrate all features supported by \textsl{rlimits} into the \textsl{cgroups} resource management. However this has not been done yet and currently remains as work in progress. 

\ifnordsec
\section*{Acknowledgment}
The authors would like to thank the anonymous reviewers for their valuable suggestions in order to improve the paper. 
\else
\fi

\bibliography{references}
\bibliographystyle{plain}

\end{document}